\documentclass[aps,pre,twocolumn,superscriptaddress,10pt,floatfix]{revtex4}
\usepackage[dvips]{graphics}
\usepackage{graphicx}
\usepackage{amsfonts}
\usepackage{amssymb}
\usepackage{amsmath}
\usepackage{subfigure}

\begin{document}

\title{Matter-wave dark solitons: stochastic vs. analytical results}
%
%

\author{S.P. Cockburn}
\affiliation{School of Mathematics and Statistics, Newcastle University,
Newcastle upon Tyne, NE1 7RU, United Kingdom}
\author{H.E. Nistazakis}
\affiliation{Department of Physics, University of Athens, Panepistimiopolis, Zografos,
Athens 15784, Greece }
\author{T.P.\ Horikis}
\affiliation{Department of Mathematics, University of Ioannina, 45110 Ioannina, Greece}
\author{
P.G.\ Kevrekidis}
\affiliation{Department of Mathematics and Statistics, University of Massachusetts,
Amherst MA 01003-4515, USA}
\author{N.P.\ Proukakis}
\affiliation{School of Mathematics and Statistics, Newcastle University,
Newcastle upon Tyne, NE1 7RU, United Kingdom}
\author{D.J.\ Frantzeskakis}
\affiliation{Department of Physics, University of Athens, Panepistimiopolis, Zografos,
Athens 15784, Greece }

\begin{abstract}
The dynamics of dark matter-wave solitons in elongated atomic
condensates are discussed at finite temperatures.
Simulations with the stochastic Gross-Pitaevskii equation reveal a noticeable,
experimentally observable spread in
individual soliton trajectories,
attributed to inherent fluctuations in both phase and density of the
underlying medium.
Averaging over a number of such trajectories (as done in experiments)
washes out such background fluctuations,
revealing a well-defined temperature-dependent temporal growth in the oscillation amplitude.
The {\it average} soliton dynamics is
well captured by the simpler
dissipative Gross-Pitaevskii equation, both numerically and via
%
an analytically-derived equation for the soliton center
based on perturbation theory for dark solitons.
%
%
\end{abstract}

\maketitle



{\it Introduction.}
Atomic Bose-Einstein condensates (BECs) constitute ideal systems for studying nonlinear
macroscopic excitations in quantum systems \cite{BEC_Book}. Excitations in the form of dark solitons and vortices
are known to arise spontaneously upon crossing the phase transition
\cite{Spontaneous_Soliton,Spontaneous_Vortex}, a feature also studied
in high-energy \cite{Kinks} and condensed-matter \cite{Condensed_Solitons} systems,
in dynamical processes \cite{Soliton_Dynamical}
and through
controlled engineering
\cite{DarkSoliton_Burger,DarkSoliton_Denschlag,DarkSoliton_Hamburg_Nature,DarkSoliton_Hamburg_PRL,DarkSoliton_Heidelberg_PRL,Vortex_Matthews_Madison}.
%
In the latter category
dark solitons are
imprinted in a controlled manner {\em after} the gas has equilibrated
\cite{DarkSoliton_Burger,DarkSoliton_Denschlag,DarkSoliton_Hamburg_Nature,DarkSoliton_Hamburg_PRL,DarkSoliton_Heidelberg_PRL}.
%
%
Although thermal effects revealed rapid soliton decay near the condensate edge \cite{DarkSoliton_Burger,Soliton_ZNG},
recent experiments at reduced temperatures ($T \ll 0.5 T_c$)
\cite{DarkSoliton_Hamburg_Nature,DarkSoliton_Hamburg_PRL,DarkSoliton_Heidelberg_PRL} found the predicted \cite{Busch_Anglin} oscillatory pattern for the averaged soliton trajectories.

To date, finite temperature dynamics of dark solitons have been 
investigated with phenomenological \cite{Phenomenological}, quasiparticle scattering
 \cite{Shlyapnikov}, and generalised mean field \cite{Soliton_ZNG} models; see also \cite{Quantum_Soliton_1,Quantum_Soliton_2} for quantum effects
in 
various background potentials. 
The former predict oscillations with increasing amplitude
(`anti-damping' \cite{Busch_Anglin}), and appear to reproduce the {\it average} soliton trajectories to 
varying degrees of accuracy, however fail to account for the {\it random} nature of the experiments.
In particular, experiments showed variations from shot to shot \cite{DarkSoliton_Hamburg_Nature,DarkSoliton_Hamburg_PRL,DarkSoliton_Heidelberg_PRL}, with
single experimental realisations revealing the existence of dark solitons
for times much longer than those for which a reproducible (or average) pattern can be generated,
an effect attributed to `preparation errors' \cite{DarkSoliton_Hamburg_Nature}.



\begin{figure}[b]
\vskip-0.4cm
\includegraphics[scale=0.25]{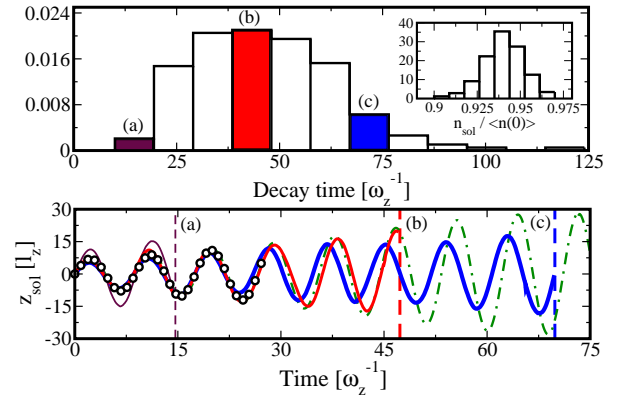}
\caption{(Color online)
Top: Normalised histograms of soliton decay times (main) and initial soliton depth,
$n_{sol}$, scaled to the average peak density $\langle n(0)\rangle$ (inset) (based on 200 realisations).
Bottom:
Individual stochastic trajectories from marked histogram bins
(for as long as they are numerically tractable),
10-realisation trajectory average (black circles) and DGPE trajectory (green, dash-dotted).
(Parameters: $N \approx 20000$ $^{87}$Rb atoms, $T=175$nK, $\omega_z=2 \pi \times 10$Hz, $\omega_\perp=2 \pi \times 2500$Hz,
$|v|=0.25c$.)
}
\end{figure}

In this Letter we show that a spread
in the trajectories of dark solitons prepared in the same manner could also arise
due to the critical dependence of {\it individual} solitons
on local phase/density fluctuations.
Modeling the soliton dynamics by
the Stochastic Gross-Pitaevskii Equation (SGPE) \cite{SGPE_Stoof,SGPE_Davis}
enables us to:
(i) obtain an {\it ab initio} calculation of the spread of individual
soliton trajectories (Fig.\ 1, top); 
(ii) 
demonstrate that although averaging over different trajectories
generates a well-defined pattern, this is restricted to times much less than the
longest observed trajectories (Fig.\ 1, bottom), consistent with experimental findings 
\cite{DarkSoliton_Hamburg_Nature,DarkSoliton_Hamburg_PRL,DarkSoliton_Heidelberg_PRL};
(iii) show that results based on stochastic trajectory averaging
can be well captured by the 
dissipative GPE (DGPE)
\cite{Phenomenological,lp,choi,ueda},
with an {\it ab initio} obtained damping coefficient; 
(iv) derive an {\it analytical} equation for the soliton center
which captures such average dynamics very well at low temperatures.

%
%

{\it Stochastic Dynamics:}
%
The 
SGPE \cite{SGPE_Stoof,SGPE_Davis} describes the condensate and lowest excitations in a unified manner, including both density and phase fluctuations,
with irreversibility and damping arising from the coupling of such modes to a thermal particle reservoir.
Assuming a `classical' approximation for the mode occupations and a thermal cloud close
to equilibrium, the SGPE reads 
\cite{SGPE_Stoof}
\begin{eqnarray}
i\hbar \partial_{t}\psi = (1-i\gamma)\left[\frac{\hbar^{2}}{2m} \partial_{z}^{2}
+ V(z) + g|\psi|^2 - \mu \right]\psi + \eta \;,
\label{sgpe}
\end{eqnarray}
where $g = 2 a \hbar \omega_{\perp}$ is the effective 1D coupling constant ($a$ is the scattering length,
$\omega_{\perp} \gg \omega_z$ the transverse harmonic confinement), and
$V(z)=(1/2)m \omega_{z}^{2} z^{2}$ the axial confining potential.
$\gamma 
= i \beta \hbar \Sigma^{\rm K}(z,t) /4$ represents the {\em ab initio} determined
dissipation arising due to the coupling to the thermal cloud ($\beta = 1/k_{\rm B}T$).
$\Sigma^{\rm K}(z,t)$ 
is the Keldysh self-energy due to incoherent collisions between condensate
and non-condensate atoms 
and $\eta$ is a noise term with gaussian correlations
$\langle \eta^*(z,t) \eta(z',t') \rangle = 2 \hbar k_B T \gamma(z,t) \delta(z-z') \delta(t-t')$  
%
%
%
%
%
(see also \cite{SGPE_Stoof,SGPE_Applications} for further details and applications to condensate properties).

Soliton experiments are modelled by first letting the system
equilibrate at a given temperature
and then introducing a dark soliton of specified velocity $v$
in the trap center by multiplying $\psi$ by
$\psi_{\rm sol} = \zeta {\rm tanh} (\zeta z / \xi) + i(v/c)$,
where $\zeta=\sqrt{1-(v/c)^2}$
($\xi$: healing length, $c$: speed of sound).
%
Although the 
soliton generation is {\it identical} in all realisations (specified by $v/c$),
fluctuations inherent in the 
atomic medium 
lead to a large variation
in the imprinted soliton:
The soliton speed $v/c=\sqrt{1-n_d/n}={\rm cos}(S/2)$ is closely related to the depth of the density minimum ($n_d$)
and the phase slip $S$ across it. As a result, fluctuations in the background density {\it upon generation} 
should modify its 
depth, whereas the speed should also 
be affected by fluctuations in the condensate phase.

The combination of these two factors leads to a slightly asymmetric spread in the initial soliton depth (Fig.\ 1, top inset), also interpreted as a stochastic change in the initial soliton speed ($\approx 30\%$ for Fig.\ 1).
Moreover, the ensuing trajectory is further modified by the local phase/density fluctuations {\it during} the SGPE evolution.
%
%
%

Soliton experiments are typically conducted in highly elongated geometries, in order to avoid dynamical instabilities \cite{Dynamical_Instabilities}.
Phase fluctuations in such geometries set in at a characteristic temperature $T_{\phi}$ \cite{Low_D_Geometries}, which can be much lower than the corresponding `critical' temperature $T_{c}$ \cite{K_vD_Tc1d}.
Although recent experiments
\cite{DarkSoliton_Hamburg_Nature,DarkSoliton_Hamburg_PRL,DarkSoliton_Heidelberg_PRL}
were conducted in the regime $T \ll T_{\phi}, T_{c}$,
where both density and phase fluctuations are largely suppressed,
soliton oscillations can still be observed in the presence of phase fluctuations  ($T \gg T_{\phi}$), provided $T \ll T_c$. To amplify the differences between individual trajectories, we thus choose realistic experimental parameters
($N \approx 20000$ $^{87}$Rb atoms, $\omega_z=2 \pi \times 10$Hz, $\omega_\perp=250 \omega_z$) corresponding to
this intermediate regime $T_{\phi} \ll T \ll T_{c}$.
This gives a phase coherence length $L_{\phi} \approx (0.1-0.25) R$ ($R$: Thomas-Fermi radius), with solitons allowed by $L_{\phi} \gg \xi$.
We focus on a relatively deep soliton which is more prone to this effect; we also anticipate
phase imprinting to further enhance differences in trajectories due to the
effect of fluctuations {\it during} the initial state preparation \cite{Quantum_Soliton_2}.


\begin{figure}[t]
\includegraphics[scale=0.21]{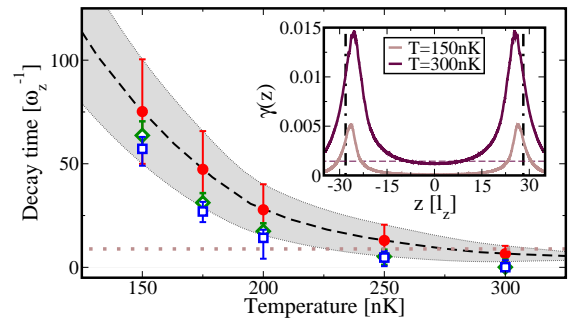}
\caption{(Color online)
Mean soliton decay times as a function of temperature obtained from the SGPE (red circles), 
or single DGPE realisations with {\em ab initio} determined $\gamma(z)$ (green diamonds),
or {\it averaged} $ \bar{\gamma}$ over $[-R/2,R/2]$ (blue squares).
Grey band indicates SGPE values within one standard deviation of the mean
decay time (for 200 runs); DGPE data shows time for soliton to decay to a
depth comparable to the average background density
fluctuations, with error bars corresponding to depths within the standard deviation of the fluctuations.
Dotted horizontal line indicates time for one oscillation.
Inset: $\gamma(z)$ (solid) for $T=150$ nK (bottom) and $300$ nK with $ \bar{\gamma}$ (horizontal) for T=300nK;
vertical lines show $R=28l_z$ (parameters as in Fig.\ 1; 
characteristic temperatures in 1D: $T_\phi = N(\hbar \omega_z)^2/k_B \mu \approx 25$ nK \cite{Low_D_Geometries}, 
$T_{c} = N\hbar \omega_z / k_B {\rm ln}(2N)\approx 900$ nK \cite{K_vD_Tc1d}).
}
\end{figure}

Typical trajectories are shown in Fig.\ 1 (bottom) up to the point where the soliton can be numerically identified over the fluctuating background,
which sets a decay time for each realisation. 
We find an asymmetric distribution of decay times, with some very long-lived trajectories.
The spread in the decay times can be best visualised via characteristic trajectories from different histogram bins (labelled (a)-(c)).
Despite their apparent differences, averaging over a sufficient number of trajectories (typically $\ge10$) washes out such sensitivity, 
generating an antidamped oscillatory pattern, with a temperature-dependent shift in both amplitude and phase (black circles). 
The average trajectory is only defined up to the earliest decay time within the set of trajectories considered (here $27 \omega_z^{-1}$), 
in 
analogy to the experimentally reproducible soliton dynamics being restricted to much shorter times than 
those of individual long-lived trajectories \cite{DarkSoliton_Hamburg_Nature,DarkSoliton_Hamburg_PRL,DarkSoliton_Heidelberg_PRL}.
The average trajectory is practically indistinguishable from an individual trajectory taken from the mean decay time bin (solid red), 
enabling us to infer the subsequent {\it average} soliton evolution from a single trajectory with a decay time close to the mean.

Fig.\ 2 shows the dependence of the soliton decay time on temperature (red circles) in the intermediate temperature range of noticeable anti-damping: At higher $T$ the soliton is lost to the fluctuating background, prior to executing one full oscillation, thus leading to a decrease in the width of the decay time histogram, and to smaller error bars in the mean decay time; although our model predicts very little damping for $T \le 100$ nK $\approx 10\% T_{\rm c}$, consistent with recent pure condensate experiments \cite{DarkSoliton_Heidelberg_PRL}, our results may overestimate the actual lifetimes, due to the neglected role of collisions in the thermal cloud \cite{Soliton_ZNG}.


%


The distribution of imprinted solitons and
decay times is the main numerical result of this paper.
Nonetheless, dynamics consistent with the average stochastic results
 can also be obtained by a simpler model discussed below.

{\it The 
DGPE and its comparison with the SGPE.}
A dissipative {\it mean-field} equation similar in form to Eq.\ (\ref{sgpe}), but without a noise term,
was first introduced in a {\em phenomenological} manner by Pitaevskii \cite{lp};
in the BEC context this was applied to damping of excitations \cite{choi},
vortex lattice growth \cite{ueda,gard} and dark soliton decay \cite{Phenomenological}.
%
A {\it numerical} advantage of the DGPE (which also restricts its predictive ability)
is that only a {\it single} realisation 
is required 
%
under the assumption that
trajectory-averaged properties should only depend on the dissipation;
this is shown in Fig.\ 2 (inset):
the self-consistent inclusion \cite{Footnote_Gamma} of the mean field potential $2g \langle | \psi|^2 \rangle$ in the expression for $\gamma$
generates a  relatively flat profile around the trap center, with peaks at the condensate edges where the thermal cloud density is greatest.
%
%
Since in the relevant dark soliton studies,
the soliton spends most of its time well within the condensate, we can extract an averaged
dissipation $\bar{\gamma}$, over a spatially-restricted region,
e.g.\ $\bar{\gamma} = \int \gamma(z) dz/R$, within
$[-R/2, \,\, R/2]$.
A simple analytical formula in the literature predicts
$\gamma(0) = \alpha (m a^2 k_B T/ \pi \hbar^2)$, with $\alpha \approx 3$ 
\cite{gard}.
We find the spatially averaged rate $\bar{\gamma}$ reveals a more pronounced scaling with temperature,
though a reasonable first estimate can be obtained in the examined temperature range using this formula with
$1/2 < \alpha < 4$.


At low temperatures, the DGPE soliton oscillations are practically indistinguishable from the SGPE ones (Fig.\ 1, bottom). 
A systematic comparison can be done 
by quantitatively comparing soliton decay times (Fig.\ 2):
in the DGPE, these are identified by
the time taken for the soliton to decay to a depth comparable to the
background density fluctuations (as predicted here by a single SGPE run, or, in general, measured experimentally).
%
We find very good agreement for both $\gamma(z)$ and $\bar{\gamma}$, within the error bars (grey bands), with
a smaller relative error at lower temperatures.
%
Since the DGPE reproduces the averaged results well in this regime
(see also Fig. \ref{fig3} below),
we now provide an {\em analytical} solution for the soliton evolution.

{\it Analytical Results.}
Upon dropping the position dependence of $\gamma(z)$ and further introducing the transformation
$t\rightarrow (1+\gamma^2)t$, the 1D DGPE takes the form:
\begin{eqnarray}
(i-\gamma) \partial_{t}\psi = \left[\frac{1}{2} \partial_{z}^{2}
+ V(z) + |\psi|^2 - \mu \right]\psi,
\label{gpe}
\end{eqnarray}
where the density $|\psi|^2$, length, time and energy are respectively measured in units of
$2a$, $a_{\perp} = \sqrt{\hbar/m \omega_{\perp}}$, $\omega_{\perp}^{-1}$ and $\hbar\omega_{\perp}$, and
$V(z)=(1/2)\Omega^{2} z^{2}$, with $\Omega = \omega_z/\omega_\perp \ll 1$. 
%
%
%
%
%
%
%
We seek a solution of Eq. (\ref{gpe}) in the form
%
%
$\psi(z,t) = \psi_b (z,t){\rm e}^{-i\theta(t)} \upsilon(z,t)$,
where $\psi_b(z,t)$ and $\theta(t)$ denote the background amplitude
and phase respectively, while
%
the dark soliton $\upsilon(z,t)$
is governed by
%
\begin{equation}
i \partial_t \upsilon +\frac{1}{2} \partial_z^2 \upsilon -
\psi_{b}^{2}  (|\upsilon|^{2}-1)\upsilon =
-\frac{ \partial_z \psi_{b}}{\psi_{b}} \partial_{z} \upsilon + \gamma \partial_t \upsilon.
\label{ups}
\end{equation}
%

We assume that the condensate dynamics
involves a fast scale of relaxation of the background to the ground state
(justified a posteriori) and that
the dark soliton subsequently evolves on top of the relaxed ground state. In the
Thomas-Fermi limit,
$\psi_{b}^{2}\approx \mu -V(z)$, and rescaling
$t \rightarrow  \mu t$, $z \rightarrow  \sqrt{\mu} z$, we obtain from Eq. (\ref{ups}) a
perturbed nonlinear Schr\"{o}dinger (NLS) equation:
%
%
%
\begin{equation}
i \partial_{t} \upsilon +\frac{1}{2}\partial_z^2 \upsilon -(|\upsilon |^{2}-1)\upsilon = P(\upsilon),
\label{pnls}
\end{equation}
where $P(\upsilon )$ stands for the total perturbation, namely,
\begin{equation}
P(\upsilon )=\frac{1}{2\mu^2} \left[ 2 \left( 1-|\upsilon |^{2} \right)V \upsilon
+ \frac{dV}{dz} \partial_z \upsilon + 2\gamma \mu \partial_t \upsilon \right],
\label{pert}
\end{equation}
%
and all terms in $P$ are assumed to be of the same order ($\gamma \sim \Omega$).
We now apply the perturbation theory for matter-wave dark solitons \cite{review}:
starting from the dark soliton solution of the unperturbed system, we seek a solution
in the form
%
%
%
$\upsilon (z,t)=\cos \varphi(t) \tanh \eta +i \sin \varphi(t)$,
where $\eta \equiv \cos \varphi(t) \left[ z-z_0(t)\right]$, and
$\varphi(t)$ and $z_0(t)$ are the slowly-varying phase ($|\varphi| \le \pi/2$)
and center of the soliton. The resulting perturbation-induced evolution equations for $\varphi$ and $z_0$, namely
$d\varphi/dt=-(1/2)\cos \varphi dV/dz
+ (2/3)\gamma \mu \cos \varphi \sin \varphi$,
and $dz_0/dt=\sin \varphi$,
%
%
%
%
%
%
lead to the following equation of motion for the soliton center,
\begin{equation}
\frac{d^{2}z_{0}}{dt^{2}} = \left[\frac{2}{3}\gamma \mu \frac{dz_{0}}{dt} - \left( \frac{\Omega}{\sqrt{2}}\right)^2 z_0 \right].
\left[1 - \left(\frac{dz_{0}}{dt}\right)^2 \right].
\label{nl_em}
\end{equation}
%
%
The nonlinear Eq.~(\ref{nl_em})
can be integrated directly to
yield the soliton trajectory:
Fig.\ 3 shows very good agreement between the prediction of Eq.\ (6) (red) and the full DGPE (black) based on the spatially integrated $\bar{\gamma}$, which are also consistent with the SGPE predictions with $\gamma(z)$.

%
%
%
In the case of a nearly black soliton (for $dz_0/dt$ sufficiently small), Eq.~(\ref{nl_em}) 
is reduced to the linearized equation $d^2 z_0/dt^2 -(2/3)\gamma \mu (d z_0/dt) + (\Omega/\sqrt{2})^2 z_0 =0$.
%
%
%
%
%
This includes the 
temperature-induced anti-damping term $\propto -\gamma dz_0/dt$,
and is reminiscent of the equation of motion derived by means of a kinetic
theory approach \cite{Shlyapnikov}.
%
%
%
For $T=0$ ($\gamma =0$) the linearized equation
recovers the constant amplitude oscillation of frequency $\Omega/\sqrt{2}$ \cite{Busch_Anglin,review}.
%
%
For $T \ne 0$ ($\gamma \ne 0$), the solutions
of the linearized Eq. (\ref{nl_em}) are
%
%
$z_0(t) \propto \exp(s_{1,2} t)$,
%
%
%
where $s_{1, 2} = \gamma\mu/3 \pm \sqrt{\Delta} (\Omega/\sqrt{2})$ are the roots of the resulting characteristic equation. 
%
%
The discriminant 
$\Delta \equiv (\gamma/\gamma_{cr})^2 -1$ (with $\gamma_{cr}=(3/\mu)(\Omega/\sqrt{2}) = 0.053$ in our units)
determine the type of motion:
%
%
%
soliton trajectories are classified into
%
%
{\it sub-critical} weak anti-damping
($\Delta<0$, $\gamma < \gamma_{cr}$),
{\it critical} ($\Delta=0$, $\gamma = \gamma_{cr}$), and
{\it super-critical} strong anti-damping ($\Delta>0$, $\gamma > \gamma_{cr}$) cases.
%
%
%
%
Assuming an initial soliton location $z_0(0)=0$ and velocity $\dot{z}_0(0)$,
the sub-critical soliton trajectory reads:
\begin{equation}
z_0(t) = \frac{\dot{z}_0(0)}{\omega_{\rm o}}{\rm e}^{\gamma \mu t/3} \cos(\omega_{\rm o} t),
\,\,
\omega_{\rm o} = \frac{\Omega}{\sqrt{2} } \sqrt{1-\frac{\gamma^2}{\gamma_{cr}^2} },
\label{sbcr}
\end{equation}
%
%
%
%
indicating an exponential increase in its maximum amplitude (Fig.\ 3 top, dashed green line), whose magnitude depends on both temperature
and chemical potential;
the oscillation frequency $\omega_{\rm o}$ is also shifted from its $T=0$ value
\cite{Soliton_ZNG}.
%
%
%
%
Corresponding trajectories
in the critical and super-critical cases
read:
$z_0(t)=\dot{z}_0(0)t\exp(\gamma \mu t/3)$ and $z_0(t)=[\dot{z}_0(0)/(s_1-s_2)][\exp(s_1 t)-\exp(s_2 t)]$.

\begin{figure}[tbp]
\includegraphics[width=8cm,height=3.5cm]{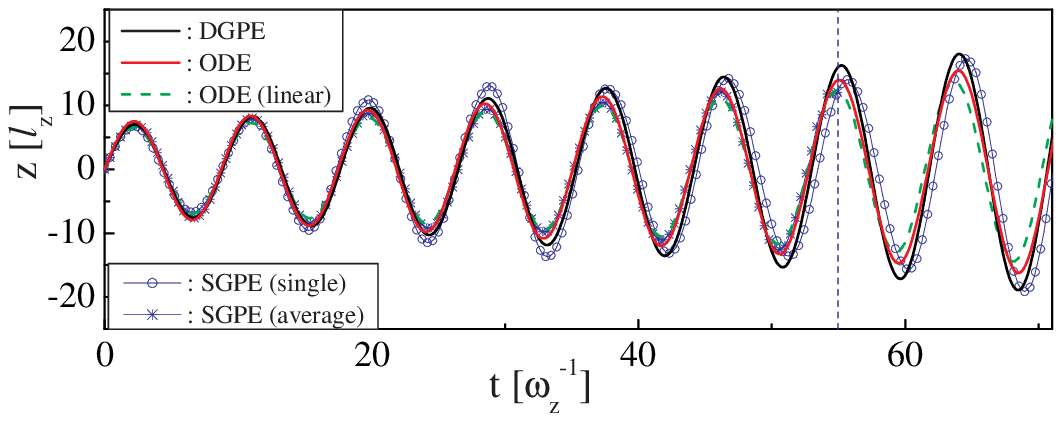}
\vskip-0.7cm
\includegraphics[width=8cm,height=3.5cm]{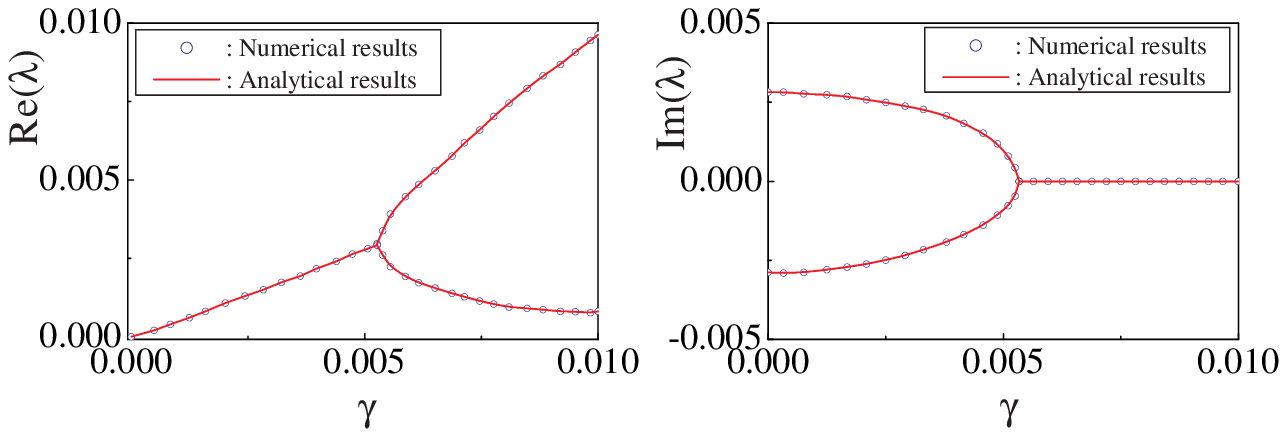}
\vskip-0.4cm
\caption{(Color online)
Top: Soliton DGPE trajectories (black) vs.\ analytical results (nonlinear: solid red; linear: dashed green) with $\gamma=\bar{\gamma}=0.00014$, and stochastic ones with $\gamma(z)$ (stars: 10-trajectory average, circles: single {\it mean-bin} trajectory).
Bottom: Dependence of the real part (left; instability growth rate) and imaginary part (right; oscillation frequency) of the unstable eigenmode of the excitation spectrum on $\gamma$: analytical results (solid red) vs.\ DGPE numerics (blue circles)
(Parameters as in Fig.\ 1, with $T=150$ nK.)
%
}
\label{fig3}
\end{figure}
The above
results are
also supported
by a linear stability analysis around
the stationary
dark soliton, $\psi_{ds}$.
This
waveform makes the right hand side of Eq. (\ref{gpe}) vanish and 
is, thus, an exact solution of the
$T \ne 0$
problem.
As rigorously proven \cite{sand}, the anomalous
(or negative Krein signature)
mode of the dark soliton
leads to an instability, upon {\it dissipative} perturbations.
In particular, the relevant mode
of the linearization around the
soliton (solution of
the eigenvalue problem arising from $\psi=\psi_{ds} + \epsilon
(\exp(\lambda t) a(x) + \exp(\lambda^{\star} t) b^{\star}(x))$ for
the eigenvalue-eigenvector pair $\{\lambda,(a,b)\}$)
%
%
acquires Re$(\lambda)>0$ for $\gamma>0$.
Fig.\ \ref{fig3} (bottom) demonstrates an {\it excellent} agreement between
the analytical prediction for the relevant eigenvalue
and the numerical
result for the excitation spectrum of the DGPE.


{\em Discussion:}
%
A full description of the rich experimental features observed in dark soliton experiments
requires a stochastic model incorporating density and phase fluctuations, in order to
model experimantally relevant shot-to-shot variations.
The stochastic Gross-Pitaevskii equation was shown
to capture these features well, leading to specific predictions for the spread of
soliton decay times with different realisations of the dynamical noise, in close analogy
to different experimental realisations.
Nonetheless,
 even within the phase-fluctuating regime,
{\it mean} soliton trajectories/decay times
are captured reasonably by the simpler dissipative
Gross-Pitaevskii equation (with additional experimental {\it or} theoretical input
required to obtain the dissipation term).
A fully
analytical solution of the dark soliton motion in excellent agreement
with the
dissipative Gross-Pitaevskii equation
applied to the finite temperature was given,
paving the way for future analytical studies of other
macroscopic excitations, such as vortices, in atomic condensates.


We acknowledge discussions with C.F. Barenghi and funding from EPSRC, NSF and S.A.R.G of U. of Athens.


\begin{thebibliography}{99}
\bibitem{Footnote_Gamma}
Replacing the mean SGPE density $\langle | \psi|^2 \rangle$ by an approximate Hartree-Fock density does not significantly affect the central region of $\gamma(z)$.
\end{thebibliography}
\end{document}